# A CQM-based approach to solving a combinatorial problem with applications in drug design.

B. Maurice Benson, Victoria M. Ingman, Abhay Agarwal, Shahar Keinan*
Polaris Quantum Biotech, 212 W Main St, Suite 200 PMB 205, Durham, NC, 27701

**Corresponding author:*
*Shahar Keinan, [skeinan@polarisqb.com](mailto:skeinan@polarisqb.com)*

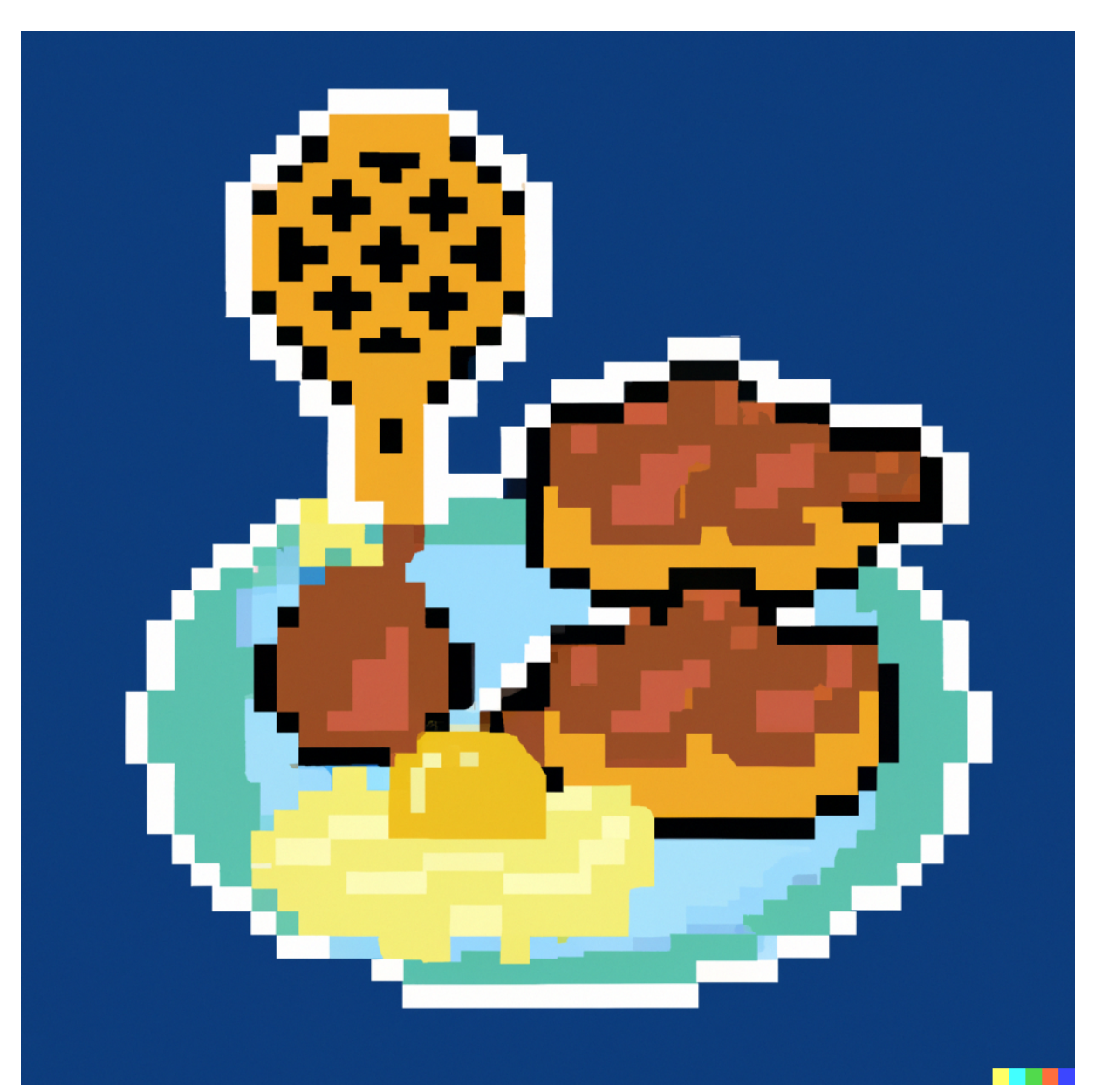

## Abstract

The use of D-Wave's Leap Hybrid solver is demonstrated here in solving a Knapsack optimization problem: finding meal combinations from a fixed menu that fit a diner's constraints. This is done by first formulating the optimization problem as a Constrained Quadratic Model (CQM) and then submitting it to a quantum annealer. We highlight here the steps needed, as well as the implemented code, and provide solutions from a Chicken & Waffle restaurant menu. Additionally, we discuss how this model may be generalized to find optimal drug molecules within a large search space with many complex, and often contradictory, structures and property constraints.

## Introduction

The goal of computational drug design is to identify a set of molecules that can interact with a target protein to produce a desired therapeutic effect. This can be formulated as an optimization problem, where the objective is to find the molecular structure(s) that optimize a given set of criteria. For example, the optimization problem could involve maximizing the binding affinity between the drug molecule and the target protein while minimizing potential side effects or toxicity. The optimization problem could also involve finding the molecular structure that satisfies certain constraints, being able to pass through the blood-brain barrier, or having a specific chemical composition. In the past, computational drug design approaches used various methods such as molecular docking, molecular dynamics simulations, and machine learning algorithms to look at possible molecular structures and identify those that best have the required properties. This was a design problem, i.e. how to change a molecule from a limited set, so it will comply with the required objectives. A different approach is to treat this as an optimization problem, i.e. design a large chemical space and use multi-objective optimization algorithms to find the set of molecules that have the required objectives. An efficient way of doing this is running the CQM algorithm on a Quantum Annealer, as we have been doing.
However, translating the structure and properties of molecules to a CQM is not a simple task. To simplify it, we have built a knapsack optimization-type “toy model”. The model implements a restaurant menu in a CQM formalism and can be used to implement, test, and evaluate the run time of such formalism.

Drug design is a subsection of a set of problems that belongs to the family of “Large Optimization Problems”. These refer to computational problems that involve finding the best solution among many possible alternatives, sometimes also called "large-scale linear programming problems" (Luo et al., 2018). These problems are typically characterized by many variables and constraints, making them computationally complex and difficult to solve using classical computing methods. Large optimization problems arise in many fields, including finance, logistics, engineering, and scientific research, and often require the use of advanced mathematical techniques and algorithms to solve (Wang et al., 2019). In the case of drug design, the solution alternatives are different molecules, and the constraints can be properties such as binding affinity, toxicity, solubility, etc. These constraints are often a balancing act among many factors or appear to be conflicting, such as requiring the drug molecule to be soluble enough for oral administration despite targeting a binding pocket in a hydrophobic region, which adds even more complexity to a classical computing optimization algorithm.

In the context of quantum computing, large optimization problems are particularly challenging, as they require the development of specialized quantum algorithms and the use of specialized hardware such as quantum annealers. The ability to solve large optimization problems is an important area of research and development in the field of quantum computing, with potential applications in a wide range of fields (Hu et al., 2020) (Bian et al., 2019).

Quantum annealers are specialized quantum computing devices that are designed to solve optimization problems (King et al., 2018). These devices use a process called quantum annealing to search for the lowest energy state of a given problem, which can be used to find the optimal solution. One of the key advantages of quantum annealers is their ability to efficiently handle large and complex optimization problems that are difficult to solve using classical computing methods (Rieffel et al., 2015). This makes them particularly useful in fields such as finance, logistics, and scientific research, where large-scale optimization problems are common. Despite their potential, quantum annealers are still relatively new and their capabilities are not yet fully understood. However, ongoing research and development in the field of quantum computing are likely to result in continued improvements and advancements in quantum annealing technology (Yarkoni et al., 2022) (Cao et al., 2019).

In this paper, we will demonstrate the use of a quantum annealer (D-Wave's Leap Hybrid solver https://www.dwavesys.com/solutions-and-products/systems/ ) to solve a "toy model" optimization problem. We will show how to formulate the specific optimization problem to a Constrained Quadratic Model (CQM) and then how to submit it to a quantum annealer.

## Methods

The knapsack optimization problem is a well-known combinatorial optimization problem that involves selecting a subset of items from a given set, subject to a weight constraint, in a way that maximizes the total value of the selected items. The problem is named after the idea of packing a knapsack with items of different weights and values (Kellerer et al., 2004).

The knapsack optimization problem has been studied extensively in classical optimization and computer science, and has also been studied in the context of quantum annealing. In the context of quantum annealing, the knapsack problem is typically formulated as a quadratic unconstrained binary optimization (QUBO) problem, which is a mathematical formulation that can be solved on quantum annealers such as D-Wave's system (Feld et al., 2019) (Awasthi et al., 2023). Various methods have been proposed for solving the knapsack problem using quantum annealers, including hybrid classical-quantum approaches that combine classical optimization techniques with quantum annealing. The knapsack optimization problem is an important benchmark problem for testing the performance of quantum annealers and is also relevant for applications in areas such as finance, logistics, and resource allocation. Here we are using a constrained version of the model that is applicable to drug design.

We have used a "Chicken & Waffle" restaurant menu as an example for the Knapsack optimization problem, with the added constraint of being able to choose a single item from each category. Here specifically, the user can choose from 4 waffle choices, 8 smears, 7 chicken choices , 7 drizzles, and 7 sides, which makes for 10,976 potential combinations. The constraints that can be applied here are varied, and here we chose to find the cheapest meal with one selection from each category and the entire meal to be less than 700 calories.

### Constrained Quadratic Model

The CQM problems are formed by trying to minimize an objective that is subject to multiple constraints
(https://www.dwavesys.com/media/rldh2ghw/14-1055a-a_hybrid_solver_for_constrained_quadr

atic_models.pdf). Equations 1-3 show how one would formulate the objective and constraints suitably for the CQM (where $x_i$ are binary variables, $a_i$ and $b_{ij}$ are real number coefficients associated with each term, and $c$ is a constant):

Equation 1, the minimized objective: $\sum_i a_i x_i + \sum_{i<j} b_{ij} x_i x_j + c$

Equation 2, the equality constraint: $\sum_i a_i x_i + \sum_{i<j} b_{ij} x_i x_j + c = 0$

Equation 3, the inequality constraint: $\sum_i a_i x_i + \sum_{i<j} b_{ij} x_i x_j + c \leq 0$

For our problem, we can think of every item on the menu as a possible decision. Since every item can either be part of the order or not, we can represent each decision as a binary variable. Going forward, we will refer to every item on the menu as a binary variable $x_i$. Our problem has 33 items on the menu, so we will have $x_1 - x_{33}$ represented in our model. A given $x_i$ can be either 0, which would represent the exclusion of the corresponding item, or 1, which would represent its inclusion.

Applying Objectives to the CQM model

The objective of the optimization is to find the cheapest meal given the constraints. For our example, we want to multiply the price of each item $p_i$ by the respective binary variable $x_i$ and add those values together. Equation 4 shows how the CQM Equation 1 is applied in this case:

Equation 4: $8x_1 + 8x_2 + 9x_3 + ... + 4x_{31} + 4.75x_{32} + 4x_{33}$

If an item is included in the order, its corresponding binary variable will be 1. Therefore, its coefficient - the price of the item - will be included in the total sum. If the item is excluded from the order, its binary variable will be 0, and therefore it will not contribute to the total price.
By setting this objective in the CQM model, the quantum annealer will attempt to return the cheapest solutions.

Applying Constraints to the CQM model

The “Chicken and Waffle” problem formulates that each meal will include only a single item from each item type: a one-hot constraint. For example, there can’t be two different waffles in the same order. This is similar to choosing a single fragment for each location in a molecular design problem. We have used an equality constraint to represent this in our model. For example, there are 4 waffle choices represented by the binary variables $x_1 – x_4$. Now we can represent that constraint by summing those binary variables and setting them equal to 1. Equation 5 shows how the CQM Equation 3 is applied in this case.

Equation 5: $x_1 + x_2 + x_3 + x_4 - 1 = 0$

Since the variables can be either 0 or 1, exactly one of the binary variables involved in the sum can be 1, and the rest will have to be zero. It's not possible for all of them to be 0 because that would equate 0 to 1, which is false. This will therefore incentivize solutions where exactly one variable in the set is chosen.

A different constraint is used when limiting the number of calories. Here, we used an inequality constraint by multiplying the calories of each item by the respective binary variable, summing the binary variables, and the inequality to less than or equal to 700. Equation 6 shows how the CQM Equation 2 is applied in this case.

Equation 6: $358x_1 + 284x_2 + 244x_3 + ... + 80x_{31} + 200x_{32} + 270x_{33} - 700 \leq 0$

## Results

In Table 1 we enumerate all possible "Chicken and Waffle" menu options, with the definition of their item type, as well as price & calories per each item. These definitions were used to build our CQM by adding the objective function (minimize the total meal price), one-hot constraints (only one item per category may be chosen), and inequality constraints (total calories must be less than or equal to 700), as described above, using the D-Wave API. The model was then submitted to the quantum annealer.

The CQM is continually sampled over a set period. The results are returned as a list of values assigned to our binary variables for each "read". Based on the CQM implementation, if a variable was unused in the solution, it will be zero; if it was used, the variable will be 1. The associated "energy" (in this case, the total price of the meal) for that sample, the number of times that same solution was seen in overall samplings of the quantum annealer, and whether all constraints were met are returned as well. The energy and number of occurrences help contextualize the optimization problem corresponding to the model. For tighter constraints, one may begin to see higher-priced meal options with a large number of occurrences, as the lower-priced regions of the feature space are now less accessible.

Table 2 shows the results of the CQM runs, specifically the best meal combinations identified by the CQM that obeys the following constraints: one item type per meal, minimal price & the total number of calories below 700.

The Supporting Information includes instructions on how to build, submit and collect the results using the D-Wave API. It also includes a git repository of the code used.

## Conclusion

In this manuscript, we have demonstrated how to turn an optimization problem into an objective and constraints; build a CQM model; and submit that model to D-Wave. We are also providing instructions on how to submit and run the same problem on D-Wave.

While our "Chicken and Waffle" problem as defined here had several possible solutions, it is possible to alter or add constraints in such a way that no "perfect" solution is possible. For example, if one were to specify that the total calories must be less than or equal to 500, honoring the inequality constraint required breaking at least one of the one-hot constraints. In this case, the quantum annealer will still return the set of sampling reads, and the constraints that were unsatisfied will be indicated. For real-world optimization problems, this presents a unique benefit: one may still consider solutions that honor as many constraints as possible while allowing for the reality that no solution exists that satisfies all constraints. A common example in drug design is when you are designing a molecule with an ideal set of properties that can never be obtained but identifying molecules that are close to those set of properties which may be “good enough”. An example may be searching for a molecule that has a low molecular weight (important for passing the blood-brain barrier) but a high surface area (important for drugs that disturb protein-protein interactions). You may not find the ideal molecules, but molecules that are close to the Pareto front are still valuable.

A simple approach to solving the Knapsack problem is to use classical computers, i.e. to enumerate and check every single possible combination and make a list of the values that the cost function yields for each. Then the combinations may be sorted by their price and calorie count to find the best solution. This method works well for small data sets but scales exponentially as the number of features and the number of variables grow.

For the problem presented here, there are 10,976 potential combinations. The minimal run time on D-Wave’s hybrid solver service is 5 seconds, which was also the total run time in our case. We would expect the same problem to run faster on a classical CPU, i.e. a shorter total run time.
On the other hand, we would expect to see a much more pronounced time difference for larger optimization problems. For example, when processing a library of a billion molecules, the total run time was about 220 CPU hours on Google Cloud Platform. However, on the D-Wave quantum annealer, using their hybrid solver service, it took a total run time of 5 seconds to obtain around 50 samples of molecules that fit the constraints (5 seconds is the minimal run time) from the same billion molecule library. If the number of molecules (available solutions) were to increase, and the number of possibilities to check grows commensurately, we can intuit that the D-Wave quantum annealer solution will scale to require substantially fewer resources than the classical CPU solution.

We have shown that the CQM can solve multi-object optimization problems on an expedited timescale. This lets us use the CQM in solving much larger optimization problems and makes the CQM a valid choice for molecular drug design.

## Supporting Information:

Git repository:

We have also included a git repo and our example dataset for this paper. In order to use the example, please download the source for this example from our GitLab repository. You can go to D-Wave to request free trial time on the QPU, add your API key to the configuration file, update the CSV with your own delicious options, and start the container!

D-Wave Instruction:

D-Wave offers an API called D-Wave Ocean that gives us tools to build the CQM and submit the model to the D-Wave QPU.
There are instructions, tutorials, and examples that can help you to formulate and submit your models.

## Tables

Table 1. All possible "Chicken and Waffle" menu options, with the definition of their item type, as well as price & calories for each option.

| name | item_type | price | calories |
|---|---|---|---|
| The Classic | waffle | $8.00 | 358 |
| Sweet Potato | waffle | $8.00 | 284.4 |
| The Gingerbread | waffle | $9.00 | 244.9 |
| The Vegan | waffle | $9.00 | 278.9 |
| Strawberry-Creme | smear | $1.75 | 70 |
| Chocolate-Hazelnut | smear | $1.75 | 72 |
| Maple-Pecan | smear | $1.75 | 165 |
| Baby-Blueberry | smear | $1.75 | 120 |
| Vanilla-Almond | smear | $1.75 | 90 |
| Orange-Honeycomb | smear | $1.75 | 70 |
| Peach-Apricot | smear | $1.75 | 70 |
| Maple Syrup | smear | $2.25 | 160 |
| Two chicken drumsticks | chicken | $7.00 | 210 |
| A chicken cutlet | chicken | $6.00 | 130 |
| Three wings | chicken | $10.00 | 430 |
| Four drumsticks | chicken | $10.00 | 420 |
| A panko-crusted chicken cutlet | chicken | $6.00 | 210 |
| Four wings | chicken | $12.00 | 573 |
| Vegetarian Cutlet | chicken | $7.00 | 270 |
| Sweet Whiskey Creme | drizzle | $2.00 | 289 |
| Honey Dijon | drizzle | $2.00 | 46 |
| Caribbean Calypso | drizzle | $2.00 | 50 |
| Asian Plum Sauce & Almonds | drizzle | $2.00 | 35 |
| Candied Pecans | drizzle | $2.00 | 132 |
| Maple Syrup | drizzle | $2.00 | 104 |
| Caramel & Salted Cashew | drizzle | $2.00 | 65 |
| A pair of eggs and creamy grits | side | $4.75 | 362 |
| Collard Greens (Spicy) | side | $4.00 | 63 |
| Smooth Grits | side | $4.00 | 182 |
| Sautéed Squash & Onions | side | $4.00 | 86 |
| Fresh-Cut Fruit | side | $4.00 | 80 |
| Southern Potato Salad | side | $4.75 | 200 |
| Mac & Cheese | side | $4.00 | 270 |

Table 2. The best meal combinations as identified by the CQM and obeying the following constraints: one item type per meal, minimal price & the total number of calories below 700.

| **Waffle** | **Smear** | **Chicken** | **Drizzle** | **Side** | **Meal Price** | **Total Calories** |
|---|---|---|---|---|---|---|
| Sweet Potato | Orange-Honeycomb | Chicken cutlet | Caribbean Calypso | Sautéed Squash & Onions | 21.75 | 620.4 |
| Sweet Potato | Peach-Apricot | Chicken cutlet | Caramel & Salted Cashew | Collard Greens (Spicy) | 21.75 | 612.4 |
| Sweet Potato | Baby-Blueberry | Chicken cutlet | Caribbean Calypso | Sautéed Squash & Onions | 21.75 | 670.4 |
| Sweet Potato | Peach-Apricot | Panko-crusted chicken cutlet | Caramel & Salted Cashew | Collard Greens (Spicy) | 21.75 | 692.4 |
| Sweet Potato | Chocolate-Hazelnut | Chicken cutlet | Caramel & Salted Cashew | Fresh-Cut Fruit | 21.75 | 631.4 |

## Competing interest declaration:

The authors are employees of Polaris Quantum Biotech Inc.